\documentclass[conference]{IEEEtran}
\IEEEoverridecommandlockouts
\usepackage{amsmath,amssymb,amsfonts,amsthm}
\usepackage{algorithmic}
\usepackage{graphicx}
\usepackage{textcomp}
\usepackage{xcolor}
\usepackage{bbold}
\usepackage{multirow}
\usepackage{balance}
\usepackage{adjustbox}

\usepackage{float}
\floatstyle{plaintop}
\restylefloat{table}

\usepackage{hyperref}
\hypersetup{%
  colorlinks=false,
  urlbordercolor=cyan,
  linkbordercolor=green
}

\newtheorem{example}{Example}

\usepackage{siunitx}

\usepackage{svg}
\usepackage{subcaption}
\usepackage{comment}
\usepackage{listings}

\usepackage[
	backend=biber,
	style=ieee,
	sortcites=true,
	url=true,
	eprint=true,
	giveninits=false,
	minnames=1,
	maxnames=3
	]{biblatex}
\usepackage{csquotes}

\renewcommand{\baselinestretch}{.96}

\def\BibTeX{{\rm B\kern-.05em{\sc i\kern-.025em b}\kern-.08em
    T\kern-.1667em\lower.7ex\hbox{E}\kern-.125emX}}

\usepackage{quantikz}

\newcommand{\calA}{\mathcal{A}}
\newcommand{\calB}{\mathcal{B}}
\newcommand{\calC}{\mathcal{C}}
\newcommand{\calD}{\mathcal{D}}

\newcommand{\calF}{\mathcal{F}}
\newcommand{\calG}{\mathcal{G}}

\newcommand{\calI}{\mathcal{I}}

\newcommand{\calL}{\mathcal{L}}
\newcommand{\calM}{\mathcal{M}}

\newcommand{\calQ}{\mathcal{Q}}
\newcommand{\calR}{\mathcal{R}}
\newcommand{\calS}{\mathcal{S}}

\newcommand{\cx}{\mathrm{CNOT}}
\newcommand{\tgate}{\mathrm{T}}

\begin{document}

\title{Lattice Surgery Compilation \\Beyond the Surface Code
\thanks{\parbox{\textwidth}{%
\IEEEauthorrefmark{1} laura.herzog@tum.de\\
\IEEEauthorrefmark{2} lucas.berent@tum.de\\
\IEEEauthorrefmark{3} a.kubica@yale.edu\\
\IEEEauthorrefmark{4} robert.wille@tum.de}}
}

\author{\IEEEauthorblockN{Laura S. Herzog\textsuperscript{1, }\IEEEauthorrefmark{1}, Lucas Berent\textsuperscript{1, }\IEEEauthorrefmark{2}, Aleksander Kubica\textsuperscript{2, }\IEEEauthorrefmark{3}, Robert Wille\textsuperscript{1, 3, 4, }\IEEEauthorrefmark{4}}
\IEEEauthorblockA{\textsuperscript{1}Chair for Design Automation, Technical University of Munich, Germany}
\IEEEauthorblockA{\textsuperscript{2}Yale Quantum Institute \& Department of Applied Physics, Yale University, New Haven, USA}
\IEEEauthorblockA{\textsuperscript{3}Munich Quantum Software Company GmbH, Garching near Munich, Germany}
\IEEEauthorblockA{\textsuperscript{4}Software Competence Center Hagenberg GmbH (SCCH), Hagenberg, Austria}
}

\maketitle

\begin{abstract}
Large-scale fault-tolerant quantum computation requires compiling logical circuits into physical operations tailored to a given architecture.
Prior work addressing this challenge has mostly focused on the surface code and lattice surgery schemes.
In this work, we broaden the scope by considering lattice surgery compilation for topological codes \emph{beyond} the surface code. 
We begin by defining a code \emph{substrate}---a blueprint for implementing topological codes and lattice surgery. 
We then abstract from the microscopic details and rephrase the compilation task as a mapping and routing problem on a macroscopic routing graph, potentially subject to substrate-specific constraints.
We explore specific substrates and codes, including the color code and the folded surface code, providing detailed microscopic constructions.
For the color code, we present numerical simulations analyzing how design choices at the microscopic and macroscopic levels affect the depth of compiled logical $\cx{}+\tgate$ circuits.
An open-source code is available on GitHub~\url{https://github.com/cda-tum/mqt-qecc}.

\end{abstract}

\begin{IEEEkeywords}
Quantum Error Correction, Fault Tolerance, Topological Codes, Compilation
\end{IEEEkeywords}

\section{Introduction}
To run large-scale quantum algorithms~\cite{dalzell2023quantum}, it is essential to incorporate quantum error correction and fault tolerance to ensure reliable results~\cite{Shor1995,Steane1996,shor1996fault,kitaev1997quantum,preskill1998reliable}.
In a fault-tolerant setting, logical qubits are encoded using a quantum error-correcting code and logical operations are designed to limit the propagation of errors.
Consequently, executing quantum algorithms on any quantum computing platform requires translating logical circuits into physical ones compatible with hardware-specific constraints---a process known as \emph{compilation}.

Most state-of-the-art research in fault-tolerant compilation~\cite{herr_lattice_2017, zhu_ecmas_2023, lao_mapping_2018, watkins_high_2024, silva_multi-qubit_2024, beverland_surface_2022, molavi_dependency-aware_2024} has focused on the surface code architecture~\cite{Kitaev_2003, Dennis_2002, fowler_surface_2012}, which can be implemented with a planar layout of qubits with nearest-neighbor connectivity.
A key challenge in this setting is the implementation of logical two-qubit gates, which may naively require non-local qubit connectivity.
A standard technique to address this issue is \emph{lattice surgery}~\cite{horsman_surface_2012, landahl_quantum_2014,litinski_game_2019}, where logical qubits are made to interact by performing measurements of additional ancilla qubits placed in the region connecting the surface code patches that encode these logical qubits.

Alternatives to the surface code can be found within the broader class of topological codes, which includes the color code~\cite{bombin_topological_2006, kubica_abc_2018} and the folded surface code~\cite{kubica_unfolding_2015, moussa_transversal_2016}.
These quantum error-correcting codes can also be implemented using a planar layout of qubits with geometrically-local (rather than strictly nearest-neighbor) connectivity.
In return, the qubit overhead associated with performing error correction may be substantially reduced.
Despite these potential gains, fault-tolerant compilation for general topological codes remained largely underexplored.

\begin{figure*}[th]
    \centering
    \includegraphics[width=0.8\linewidth]{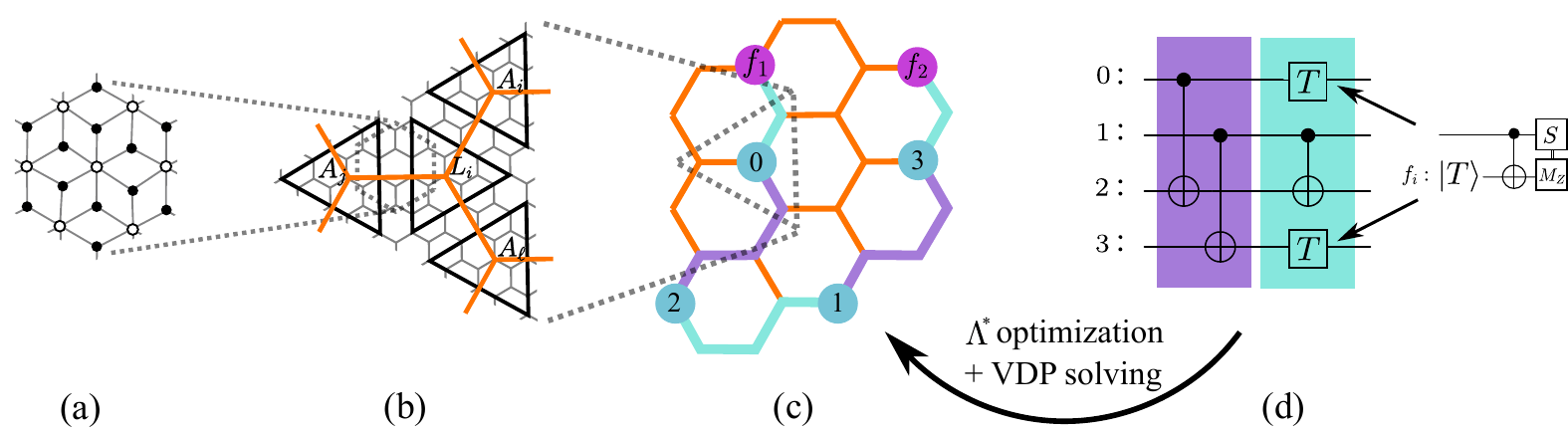}%
    \caption{Lattice surgery compilation overview. 
    (a) Geometrically local physical qubit connectivity for physical data (black circles) and ancilla (white circles) qubits. 
    (b) Substrate $\calS$ (grey) for the color code, together with patches forming logical ancilla and data qubits (magic state patches not displayed in this cutout). 
    This is referred to as the \emph{microscopic} level.
    (c) Zooming out yields the \emph{macroscopic} level with the routing graph $\calR$ (orange) and specific allocation of logical data patches, represented by light blue nodes with integer labels.
    Magic state patches are displayed as pink nodes $f_i$.
    Furthermore, the colored paths represent a possible routing for the exemplary input circuit in (d), where parallel executable gates are displayed in the same color.%
    \label{fig:fig1}}
\end{figure*}

In this work, we address the problem of fault-tolerant compilation for topological codes by introducing a general formalism that is applicable \emph{beyond} the surface code. This spans multiple levels of abstraction as illustrated in~\autoref{fig:fig1}.
On the microscopic level, we define the concept of a \emph{code substrate}---a blueprint for realizing a topological code and its corresponding lattice surgery scheme; see~\autoref{fig:fig1}{b}.
We present two concrete examples of such substrates that support the color code and the folded surface code.
While the folded surface code requires a substrate with slightly more complex qubit connectivity than the color code, it is expected to achieve higher fault-tolerant thresholds while retaining key advantages of the color code, such as transversal single-qubit Clifford gates.

At the macroscopic level, we abstract from the microscopic details by allocating logical qubits to regions of the substrate and constructing a coarse-grained \emph{routing graph} as shown in~\autoref{fig:fig1}{c}.
This abstraction enables us to design a method for routing lattice surgery operations to implement two-qubit logical gates.
We assess our approach through numerical simulations of logical $\cx{}+\tgate$ circuits using the color code substrate.
Our results highlight how logical qubit allocation influences routing efficiency, and how $\tgate$ gates and logical circuit parallelism impact the overall compilation.
More broadly, our work establishes foundational methods for compiling logical circuits in lattice-surgery-based architectures with topological codes, marking a first step toward practical fault-tolerant compilation beyond the surface code.

\section{Main Ideas and Problem formulation}\label{sec-problem-formulation}

In this work, we consider the problem of compiling logical circuits consisting of Clifford$+\tgate$ gates to lattice surgery operations assuming that $\tgate$ gates are realized using magic state distillation and injection~\cite{bravyi_universal_2005, lee_low-overhead_2024}. 
However, we do not assume any specific distillation procedure.

Previous work on fault-tolerant computation such as~\mbox{\cite{beverland_surface_2022, zhu_ecmas_2023, molavi_dependency-aware_2024, lao_mapping_2018, watkins_high_2024, silva_multi-qubit_2024}}, focused on the surface code and assumed a rectangular logical qubit connectivity. %
To consider the compilation problem more generally, i.e.,  without assuming any particular qubit connectivity and quantum error-correcting code beforehand, we distinguish the \emph{microscopic} and \emph{macroscopic} levels of abstraction.
At the microscopic level, the main problem is how to realize a lattice surgery scheme that implements logical two-qubit operations. 
At the macroscopic level, the central task is to assign logical qubits to physical locations, obtaining a logical routing graph whose shape is dictated by concrete choices at the microscopic level. 
Then valid paths between logical data qubits have to be identified to implement the logical quantum circuit.

To showcase the resulting formalism, consider the example of compilation with the 2D color code illustrated in~\autoref{fig:fig1}. 
We start from the assumption of some geometrically local connectivity of physical data qubits (black circles) and ancilla qubits (white circles) as illustrated by the graph in~\autoref{fig:fig1}{a}. 
In particular, we assume a stacked planar architecture that consists of a constant number of planar layers with qubit connectivity that is local within each layer and between layers (but not necessarily nearest-neighbor). Depending on the hardware it may be favorable to use indeed a stacked planar architecture or to project the stacked layers into one while accepting a more complex qubit connectivity.

Next we define a code \emph{substrate} that captures the essential microscopic details of the hardware architecture.
The substrate is a blueprint for realizing quantum error correction with some topological quantum code. %
For instance, a color code substrate is depicted by the gray regular hexagonal tiling in~\autoref{fig:fig1}{b}. 
In this example, data qubits and stabilizers of the code correspond to, respectively, the vertices and faces of the tiling.

Building on this, we split the substrate into several regions (black triangles in~\autoref{fig:fig1}{b}) that are dedicated to support logical data qubits, magic state patches, and the remaining \emph{routing space}. 
For concreteness, we assume that the magic states are encoded in the same code as the logical qubits. %
Moreover, it is instructive to think of the routing space as being further subdivided into several ancilla patches.
A logical data patch $L_i$ and three ancilla patches are illustrated in~\autoref{fig:fig1}{b}.
The ancilla regions form the building blocks of what we refer to as \emph{snakes} on the substrate. They are ancilla regions that are used to implement lattice surgery schemes between logical data patches.
Note that a logical patch may also correspond to a logical ancilla qubit placed in the routing space.

A concrete assignment of logical data patches, magic state patches, and ancilla regions on the substrate determines a logical routing graph as shown in~\autoref{fig:fig1}{c}. Its vertices correspond to logical data patches, magic state patches, and ancilla patches of the routing space.
The edges of the routing graph indicate the connectivity between the regions on the substrate. 
Paths on this macroscopic routing graph correspond to microscopically defined snakes together with a logical ancilla patch that are used to perform $\cx$ gates between logical data patches.

Overall, given a logical quantum circuit, for instance the one shown in~\autoref{fig:fig1}{d}, the macroscopic compilation problem consists of two main tasks that we formulate as combinatorial optimization problems: i) assigning logical qubit labels to logical data vertices of the routing graph and ii) finding non-overlapping paths on the routing graph %
such that the compiled circuit depth is minimized. 
For instance, integer labels in~\autoref{fig:fig1}{d} are mapped onto the logical routing graph as displayed by the corresponding labels on vertices in~\autoref{fig:fig1}{c}.

Having the main ideas established we present the basics of quantum error-correction and lattice surgery in~\autoref{sec-background}. 
This is followed by~\autoref{sec:overview} in which the general formalism for lattice surgery compilation is presented.
A solution for the macroscopic compilation task is presented in~\autoref{sec:compilation-hex}. 
We detail the two considered substrates in~\autoref{sec-substrates}, accompanied by numerical simulations for the macroscopic compilation of the color code substrate in~\autoref{sec-numerics}, before concluding in~\autoref{sec-conclusion}.

\section{Basics of Quantum Error Correction and Lattice Surgery
}\label{sec-background}

\subsection{Topological Quantum Codes}
\begin{figure}[t]
    \centering
    \begin{subfigure}[b]{0.3\linewidth}
        \centering
        \includegraphics[width=\linewidth]{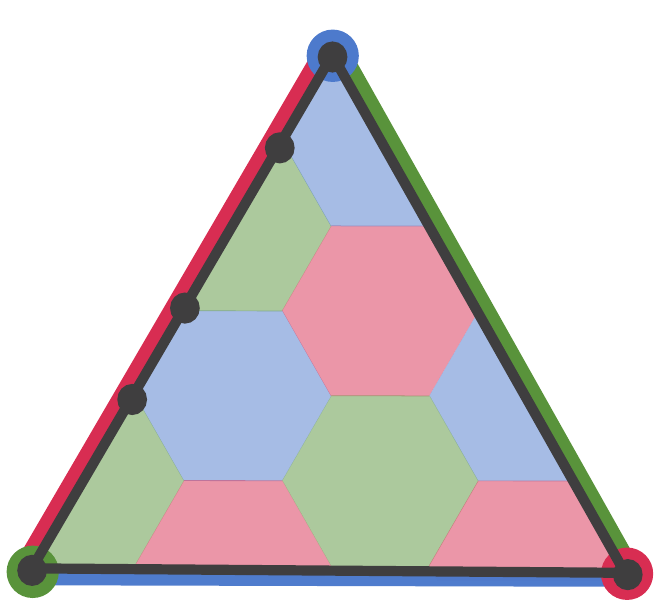}
        \caption{}
        \label{fig:subfig_a}
    \end{subfigure}
    \hfill
    \begin{subfigure}[b]{0.3\linewidth}
        \centering
        \includegraphics[width=\linewidth]{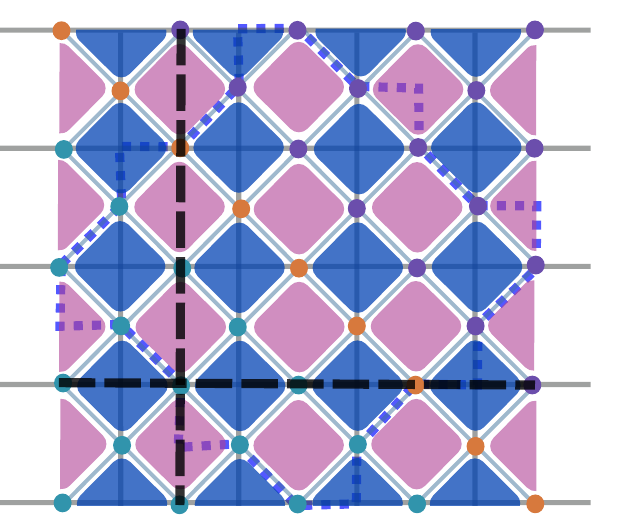}%
        \caption{}
        \label{fig:subfig_b}
    \end{subfigure}
    \hfill
    \begin{subfigure}[b]{0.3\linewidth}
        \centering
        \includegraphics[width=\linewidth]{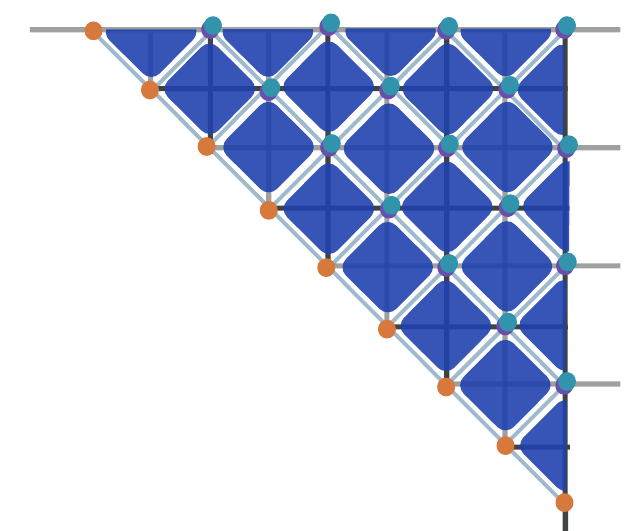}
        \caption{}
        \label{fig:subfig_c}
    \end{subfigure}
    \caption{(a) $d=5$ color code on a hexagonal tiling with data qubits on the vertices. 
    Each face defines both a $X$ and $Z$ stabilizer on the incident qubits. 
    Both $X_L$ and $Z_L$ logical operators are supported on the boundaries of the triangular region. 
    (b) $d=5$ surface code, where $Z$ stabilizers are pink faces and $X$ stabilizers are blue faces. 
    The dotted line encloses a region supporting the rotated surface code, where faces only partly contained in the boundary correspond to weight 2 stabilizers. 
    (c) Folded surface code obtained by folding along the orange qubits on the diagonal. 
    This procedure stacks $X$ faces on $Z$ faces and vice versa.
    }
    \label{fig:combined}
\end{figure}

\emph{Topological codes}~\cite{bombin_topological_2006, bombin_introduction_2013, bombin_topological_2010}
are a class of quantum error-correcting codes that are defined by placing qubits on some manifold; we restrict our attention to stabilizer codes~\cite{gottesman1998heisenberg}.
In particular, in two dimensions checks and data qubits are often associated with faces and vertices of some tiling.
Each check of the stabilizer group $S$ is local, supported only on a constant number of qubits,
whereas logical qubits are encoded non-locally.
Prominent examples for topological codes are color codes and surface codes that are defined using different tilings of 2D surfaces. 

Two-dimensional \emph{color codes}~\cite{bombin_topological_2006} are defined on 3-colorable and 3-valent tilings. 
An important family of color codes are triangular color codes on a hexagonal tiling whose faces and boundaries are three-colored.
An example is displayed in~\autoref{fig:subfig_a}.
Here, each face contained within the triangular region defines both an $X$ and a $Z$ check. 
$Z$ and $X$ logical Pauli  string operators~\cite{bombin_topological_2006} 
terminate at boundaries of the same color as displayed in the figure, or connect three boundaries of different colors. 
The code \emph{distance} $d$ corresponds to the minimum weight of a non-trivial logical operator. 
In the example in~\autoref{fig:subfig_a} the distance is $d=5$.

The \emph{surface code}~\cite{Kitaev_2003, Dennis_2002, fowler_surface_2012} can be defined on a rectangular tiling with two different types of boundaries.
The checks are placed in a checkerboard pattern. 
An example is shown in~\autoref{fig:subfig_b}. 
Here, representatives of logical operators can be defined as strings that connect boundaries of the same type. 
In this example, horizontal strings correspond to $Z_L$ (connecting so-called rough boundaries) and vertical strings (connecting so-called smooth boundaries) to $X_L$. 
Therefore, the distance of the displayed example is $d=5$. 
We note that a rotated variant of the surface code, which is supported within a region enclosed by the dotted line in~\autoref{fig:subfig_b}, leads to a smaller qubit overhead.

In its original formulation, the surface code does not admit transversal implementation of the logical $\mathrm{S}$ gates that are compatible with geometrically-local qubit connectivity.
This issue can be circumvented by folding~\cite{moussa_transversal_2016,kubica_unfolding_2015} the surface code along the diagonal (orange nodes in~\autoref{fig:subfig_b} and~\autoref{fig:subfig_c}) such that faces of different check type are stacked on top of each other.
Additionally, smooth and rough boundaries are also stacked, such that each boundary enables access to both logical $X_L$ and $Z_L$ representatives, similarly as for the color code.

\subsection{Lattice Surgery}\label{sec:bg-ls}
\begin{figure}[t]
    \centering
    \begin{subfigure}[b]{0.5\linewidth}
        \centering
        \resizebox{\linewidth}{!}{
            \begin{quantikz}
                \lstick{$\ket{c}_L$}       & \qw        & \qw              & \gate[2]{M_{ZZ}}  & \gate{Z^{a+c}} & \qw &\qw \\
                \lstick{$\ket{0}_L$}       & \gate[2]{M_{XX}} & \qw        & \qw   & \gate{M_X}      & \gate{Z^c} &  \\
                \lstick{$\ket{t}_L$}       & \qw        & \qw              & \qw            &\gate{X^b} & \qw &\qw
            \end{quantikz}
        }
        \caption{}
        \label{fig:subfig_a_CNOT}
    \end{subfigure}
    \hfill
    \begin{subfigure}[b]{0.4\linewidth}
        \centering
        \resizebox{\linewidth}{!}{
            \begin{quantikz}[]
                \lstick{$\ket{\psi}_L$} & \ctrl{1} & \qw &\gate{S} &\qw\\
                \lstick{$\ket{T}_L$} & \targ{} & \qw &\gate{M_Z} \wire[u]{c}
            \end{quantikz}
        }
        \caption{}
        \label{fig:subfig_b_TGATE}
    \end{subfigure}
    \caption{(a) Measurement-based representation of the $\cx{}$ gate. 
    $a,b,c$ are the measurement outcomes of $M_{XX}, M_{ZZ}, M_{X}$, respectively. 
    (b) Injection of a logical magic state on a logical qubit with a $\cx{}$ gate.}
    \label{fig:LS}
\end{figure}
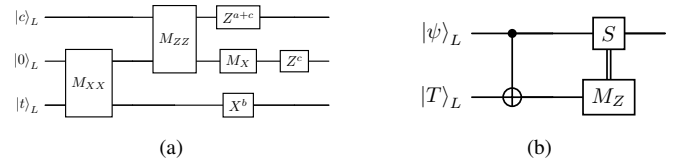

Quantum computation requires a universal gate set, such as $\{\cx, \mathrm{H}, \tgate\}$. 
In general, transversal implementation of the two-qubit $\cx{}$ gate requires a non-local qubit connectivity. %
Therefore, to implement $\cx{}$ gates with local connectivity, one may resort to \emph{lattice surgery}~\cite{horsman_surface_2012}. 

To perform a $\cx{}$ gate with lattice surgery operations, one decomposes it in a measurement-based scheme~\cite{gottesman1998fault,horsman_surface_2012, landahl_quantum_2014} on the logical level as shown in~\autoref{fig:subfig_a_CNOT}.
The challenging parts are the joint logical $M_{XX}$ and $M_{ZZ}$ measurements~\cite{landahl_quantum_2014}.
To perform $M_{ZZ}$, we first initialize physical ancilla qubits between logical (data/ancilla) patches in $\ket{+}$. 
Then we \emph{merge} both logical qubits by measuring joint stabilizers, thereby defining a code with support on the physical ancillas and the original logical patches. 
Moreover one needs to choose a subset of stabilizers whose measurement outcomes are equivalent to the value of the operator $Z_LZ_L$ on the separate logical patches, such that the desired measurement result can be retrieved. 
Finally, after performing error-correction, the merged object is \emph{split} again by measuring the physical ancilla qubits destructively in the $X$ basis.
This is again followed by error-correction for the separate logical qubits. 
For $M_{XX}$ the merged code may be adapted accordingly, e.g., physical ancillas are initialized in $\ket{0}$ and measured in the $Z$ basis. Note that there are also lattice surgery schemes without physical ancillas between the patches~\cite{landahl_quantum_2014}, which can simplify above procedure.

In the context of our work it is thus important to construct snakes between distant logical (data/ancilla) patches in a way such that the snake together with the logical qubits can be merged to form a joint code.
Furthermore, one has to guarantee that among the joint stabilizers one can combine measurement results such that the desired logical outcome of $Z_LZ_L/X_LX_L$ can be inferred.
Finally, there needs to be enough space between logical data qubits such that we can define a logical ancilla patch that is initialized in $\ket 0$ and used in lattice-surgery implementation of a $\cx{}$ gate (\autoref{fig:subfig_a_CNOT}).

The microscopic details of the lattice surgery schemes that are specific to the two examples of color codes and folded surface codes will be detailed in~\autoref{sec-cc-snakes} and~\autoref{sec:fsc-snakes}.

In order to complete the universal gate set, it is required to perform $\tgate$ gates. 
This can be done by using the logical circuit in~\autoref{fig:subfig_b_TGATE}, where the $\cx{}$ gate is performed as explained above. 
$\tgate$ magic states are considered to be the outputs of specified magic state factory patches.
We do not analyze in detail the inner workings of the factories; rather, we assume that each of them produces a $\tgate$ state after some fixed reset time.

\subsection{Related Work}
Let us conclude this section by reviewing how the resulting macroscopic compilation task is solved in related work.
Surface code compilation with a universal or smaller gate set was explored in~\cite{herr_lattice_2017, zhu_ecmas_2023, lao_mapping_2018}.
Furthermore, it has been shown that the complexity of optimizing a quantum circuit in the lattice surgery model is NP-hard~\cite{herr_optimization_2017}. 
Related work by Litinski~\cite{litinski_game_2019} motivated further research~\cite{watkins_high_2024, silva_multi-qubit_2024}, though it was argued that commuting
Clifford gates until the end of the circuit may limit the parallelism~\cite{beverland_surface_2022, molavi_dependency-aware_2024}. Whether or not Clifford gates are commuted to the end may depend on the circuit being compiled~\cite{leblond2025quantumresourcecomparisonleading}.
The aforementioned works focus on surface codes only; consequently, fault-tolerant compilation for other codes remained underexplored, despite some early ideas~\cite{litinski_combining_2017}.

Using color codes instead of surface codes offers several advantages.
For instance, the color code has a higher qubit density, supports transversal Clifford gates and hosts both $X_L$ and $Z_L$ per boundary. This simplifies the routing to some extent by removing the constraint of connecting a specific choice of boundaries. 
However, decoding algorithms for the color code tend to perform worse~\cite{Kubica2023,chamberland_triangular_2020,sahay2022decoder,gidney2023new,lee_color_2025}.
Nevertheless, recent experiments with color codes~\cite{postler2022demonstration,ryan2022implementing,rodriguez2024experimental,pogorelov2025experimental} and importantly, lattice surgery demonstration~\cite{lacroix_scaling_2024} underpin the potential of color codes as a viable fault-tolerant architecture.

Concerning the surface code, permitting a stacked planar architecture or relaxing the constraint on nearest-neighbor qubit connectivity enables to use the folded surface code~\cite{kubica_unfolding_2015,moussa_transversal_2016}.
It retains many advantages of the color code while allowing straightforward decoding, the same way as for the standard surface code~\cite{Dennis_2002,higgott_sparse_2023}. 
However, to the best of our knowledge, no previous work on compilation with these codes exists.

\section{Formalism for Fault-Tolerant Compilation}\label{sec:overview}
\begin{figure}[t]
    \centering
    \begin{subfigure}[b]{0.45\linewidth}
        \centering
        \includegraphics[width=\linewidth]{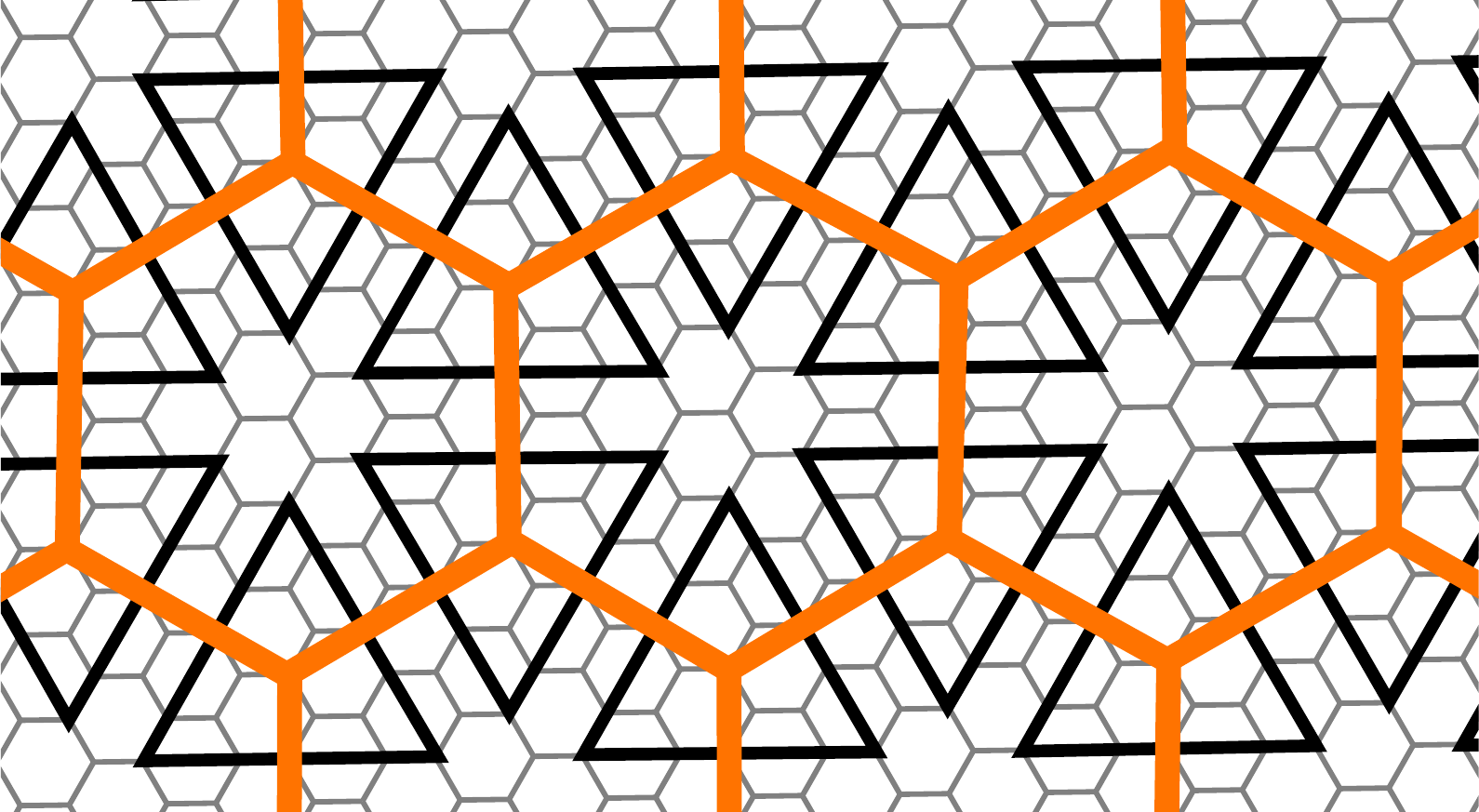}
        \caption{}
        \label{fig:substrate_subfiga}
    \end{subfigure}
    \hfill
    \begin{subfigure}[b]{0.45\linewidth}
        \centering
        \includegraphics[width=\linewidth]{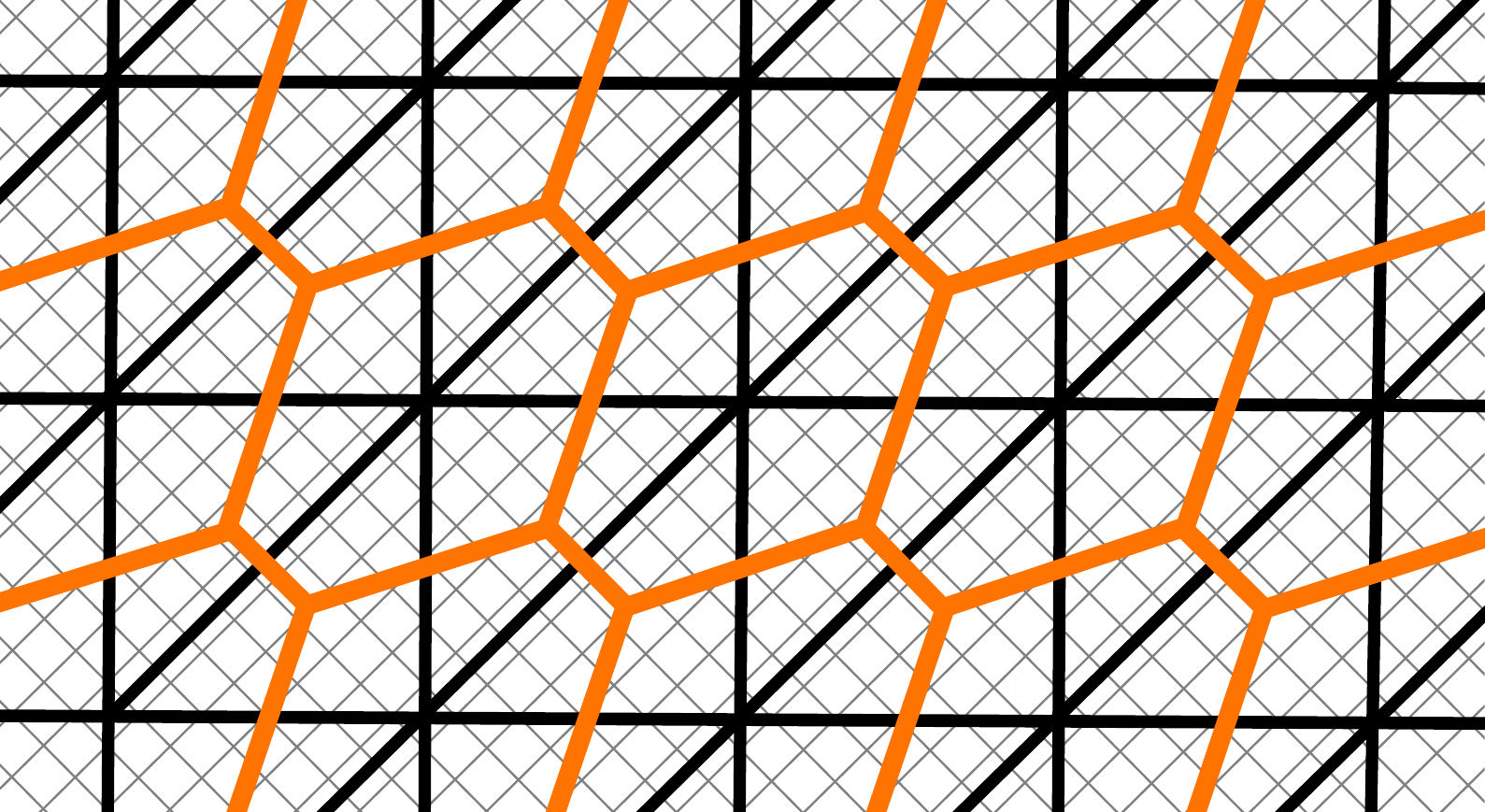}
        \caption{}
        \label{fig:substrate_subfigb}
    \end{subfigure}

    \caption{Code substrate $\calS$ (depicted in grey) for (a) the color code and (b) the surface code. 
    The logical patches from $\calL, \calA, \calF$ are bounded by thick black lines. 
    Unlike the ancilla patches, the data patches  on the surface code substrate are folded, which is not explicitly displayed.  
    The resulting routing graph $\calR$ is the orange hexagonal graph, which is the same for both substrates. 
    Both depicted substrates have $d=5$ logical patches.}
    \label{fig:overview-figure}
\end{figure}

To outline the general formalism of the considered compilation task, we define microscopic details about the code structure and derive the macroscopic picture, where compilation can be viewed as a mapping and routing task.

On the physical level we assume that the local qubit connectivity is given by a graph (cf.~\autoref{fig:fig1}{a}) that may have a constant number of layers of physical qubits such that qubits within each layer and between neighbouring layers can have geometrically local %
interactions.
Note that such an architecture is less hardware-demanding than the one needed to implement bivariate bicycle codes~\cite{bravyi_high-threshold_2024,kovalev2013quantum}; as such, it may be easier to realize with superconducting qubits~\cite{Devoret2013,Blais2021}, as well as neutral atoms~\cite{Saffman2010,Browaeys2020} or trapped ions~\cite{Cirac1995,Leibfried2003}.
Physical data and ancilla qubits (black and white in~\autoref{fig:fig1}) are placed on the vertices and their connectivity is given by the edges. 
This is the basis for choosing the code \emph{substrate}, which is a tiling $\calS$ %
(\autoref{fig:fig1}{b}) that can be used to realize logical qubits encoded in some topological quantum code. 
The substrate allows us to view a concrete physical qubit connectivity abstractly.
We use the substrate to illustrate the layout of data qubits and geometrically-local checks of the code (faces of $\calS$).

We distinguish three sets $\calL, \calA, \calF$ of disjoint regions of vertices (\autoref{fig:fig1}{b}) of the substrate.
The regions in $\calL = \{L_1, L_2,...,L_\ell\}$ correspond to logical data patches and $\calF = \{F_1, F_2,...,F_f\}$ to magic state patches, where magic state factories output magic states. 
The set $\calA$ partitions the remaining region of the substrate into ancilla patches $\calA^* = \{A_1,A_2,...,A_a\}$ and additional physical ancilla qubits
\mbox{$\calA\setminus \calA^*$}. 
The latter may be necessary to microscopically connect patches, depending on the substrate. 
A logical qubit can also require less space than the full, predefined patch provides. 
Both considered substrates are displayed in~\autoref{fig:overview-figure}.

Finally, a choice of a topological quantum code and a substrate together with the sets $\calL,\calA,$ and $\calF$ determines a logical \emph{routing graph} $\calR=(V_\calR, E_\calR)$ (\autoref{fig:fig1}{c}) that has a vertex for each set in $\calL \cup \calA^* \cup \calF$, i.e.,  $V_\calR = V_{\calL} \cup V_\calA \cup V_\calF$.
Its edges, $E_\calR$ determine the connectivity between logical data, ancilla patches, and magic state patches as illustrated in~\autoref{fig:fig1}{c}.

In this work we assume that the input circuit $\calC$ is a logical quantum circuit that contains logical $\cx+\tgate$ gates. 
To obtain a universal gate set, single-qubit Clifford gates would be necessary as well, but they are not considered here since the computational hardness of the routing challenge for topological codes is dominated by multi-qubit operations~\cite{beverland_surface_2022, molavi_dependency-aware_2024, silva_multi-qubit_2024, zhu_ecmas_2023}.
Moreover, our concrete choices of codes and substrates detailed below allow for transversal single-qubit Clifford operations.

To realize logical two-qubit
operations we use lattice surgery schemes between logical patches on the substrate.
Upon a choice of $\calL, \calA, \calF$ and a code, lattice surgery requires the implementation of specific stabilizer measurements to connect logical patches as reviewed in~\autoref{sec:bg-ls}.
To this end we realize what we refer to as \emph{snakes} on the substrate. 
Snakes consist of neighboring ancilla regions and allow for the merge of distant logical (data/ancilla) patches to measure joint two-qubit logical Pauli operators.
Throughout we assume that paths on $\calR$ between distant logical data qubits contain both a snake as well as a logical ancilla patch, together with potentially needed physical ancilla qubits from $\calA\backslash\calA^*$ to perform a logical $\cx$ gate.

The concrete design of the stabilizers of a snake requires special care as the protocol must be implemented in a fault-tolerant and distance-preserving way.
We detail two concrete constructions for logical color code patches and color code snakes, and folded surface code patches and surface code snakes in~\autoref{sec-cc-snakes} and~\autoref{sec:fsc-snakes}, respectively.

An important optimization task of the macroscopic compilation problem is to place the logical qubit patches in such a way that the paths between interacting logical data and magic state patches can be routed efficiently. 
The potential for efficiency is determined by the inherent parallelism of the logical quantum circuit, which corresponds to how many gates can be applied simultaneously, i.e., the depth of the circuit.
This amounts to defining an injective labeling $\Lambda : \calL \to \mathbb{N}$. 
Since each logical patch $L_i\in \calL$ corresponds to a vertex in $V_\calL$, the map $\Lambda$ induces a labeling of the routing graph vertices $\Lambda^{*}:V_\calL \to \mathbb{N}$ with $\Lambda(L_i) = \Lambda^{*}(r)$, where $r$ is the vertex corresponding to the patch $L_i$.

\begin{example}
    The circuit ~\autoref{fig:fig1}{d} has two layers (violet and teal) where gates inside a layer can in principle be executed in parallel. 
    (c) shows a routing solution with some choice of $\calL, \calA$, and $\calF$ for the color code with hexagonal routing graph $\calR$.
    Furthermore, the values of a labeling $\Lambda^*$ are depicted by the integer vertex labels. 
    In this example, the compiled circuit has the same depth as the initial circuit.
\end{example}

The macroscopic compilation depends on $\calR$ and possible additional constraints that define \emph{valid} paths on the routing graph, which in turn depend on the concrete choice of substrate.

\section{Macroscopic Compilation of $\cx + \tgate$ circuits}\label{sec:compilation-hex}\label{sec-compilation}
\begin{figure}[t]
    \centering
    \includegraphics[width=0.9\linewidth]{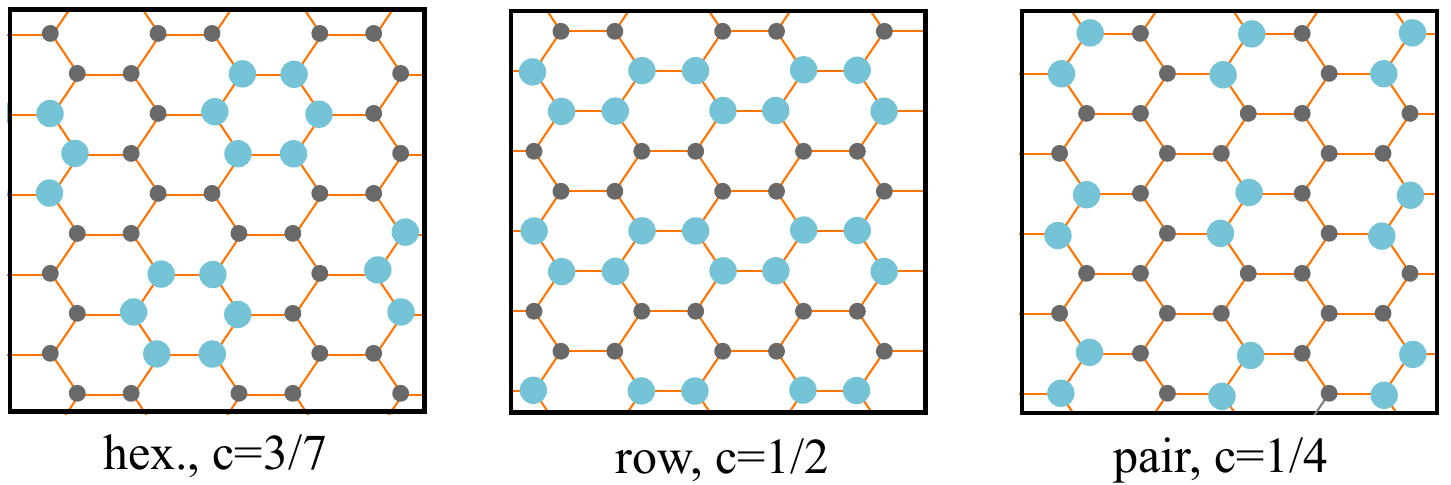}
    \caption{Considered layouts, i.e., a choice of $\calL$ and $\calA$ that determines the routing graph $\calR$, and their asymptotic packing ratio $c$, which we define as the number of logical data qubit patches (assuming that the folded surface code forms one patch) to the total number of patches. 
    Light blue and gray nodes depict logical data qubits and ancilla patches, respectively.
    }
    \label{fig:layouts_nx}
\end{figure}
The logical routing graph $\calR$ is defined by specifying the substrate $\calS$ and the regions $\calL, \calA$, and $\calF$ (corresponding to logical data patches, ancilla patches, and magic state patches).
On this macroscopic level, one has to find non-overlapping paths or a \emph{routing} on the graph between logical data patches and magic state patches according to the logical input circuit with the objective to reduce the depth of the final schedule of lattice surgery operations.
Since the assignment of logical patches and factories influences the routing problem, the mapping $\Lambda^*$ should be optimized to exploit as much inherent parallelism of the circuit as possible.

As displayed in \autoref{fig:overview-figure}, both substrates considered in this work abstract to a hexagonal routing graph. 
A specific choice of $\calL$ and $\calA$ induces a \emph{layout} of logical data and ancilla patches on the graph. 
Some possible choices for such layouts are displayed in~\autoref{fig:layouts_nx}, where elements of $V_\calL$ are shown in light blue and elements of $V_\calA$ as smaller, grey vertices. Note that layout choices depend on the considered substrate, since a layout must allow us to find valid paths between all elements of $V_\calL$ in principle. 
For the color code substrate a valid path is any path on the graph $\calR$ that is not a single edge, between two logical data patches. 
Such paths could not host the logical ancilla patch needed for a $\cx$ and are thus not valid. 
Since each patch on the color code substrate has both $X_L$ and $Z_L$ available on each boundary, no further restrictions have to be drawn into consideration. 
For the surface code substrate with folded surface codes as logical data patches, however, the availability of logical operators for lattice surgery along the boundaries of the (ancilla) patches is limited and more rules for the valid paths must be defined. 
This is detailed in~\autoref{sec:fsc-snakes}.
Thus all three layouts in~\autoref{fig:overview-figure} can be considered for the color code substrate, while for the surface code substrate, only the pair layout allows to define valid paths between any pair of logical patches.
For both substrates the magic state patches are placed on the boundary of the layout such that the magic state factory is left unspecified outside of the chosen architecture and is assumed to have enough space.

In the following sections we detail how we can solve the routing and mapping problem for the macroscopic compilation, similar to~\cite{beverland_surface_2022, molavi_dependency-aware_2024, zhu_ecmas_2023, watkins_high_2024}.

\subsection{Shortest-First VDP Solving}\label{sec:shortest-first-vdp}
Let us start by considering logical circuits with $\cx{}$ gates between disjoint pairs of qubits only.
Given a routing graph $\calR = (V_\calR, E_\calR)$ and a labeling of logical data patches $\Lambda^*$, one can formalize the search for paths as \emph{vertex disjoint path problem} (VDP)~\cite{chuzhoy_approximating_2015, chuzhoy_improved_2016, bacco_shortest_2014}. 
The aim of the VDP is to find non-overlapping paths between pairs of vertices $C=\{(r_{1}, r'_{1}),...,(r_k, r'_k) \mid r_i \neq r_j, r'_i \neq r'_j, r_i \neq r'_j\, \forall i,j\} \subset V_\calL \times V_\calL$. 
The considered vertices in $V_\calL$ are mapped to logical qubit labels via a choice of $\Lambda^*$. 
Hence, the pairs to connect---i.e., the elements of $C$---represent the control and target qubits of $\cx{}$ gates and the paths contain the snakes and logical ancillas to perform lattice surgery between pairs of logical data qubits. 
Overall, $C$ can be considered as quantum circuit consisting of $\cx{}$ gates. 

A simple greedy algorithm~\cite{chuzhoy_approximating_2015} to find solutions for the VDP is to start with an empty solution $W_i$ that is iteratively filled with paths $p_j$. 
We start by computing the shortest path with Dijkstra's algorithm between each pair in $C$, choose the shortest among those paths and add it to the solution
$W_i = W_i \cup \text{argmin}_{p_i \in \{\textsc{DIJKSTRA}(r_j, r'_j) \mid (r_j, r'_j) \in C\}} |p_i|$.
Then the vertices in the chosen path $p_i$ are removed from $\calR$. 
As pointed out in the beginning of the section, a substrate may require nontrivial constraints on  paths on $\calR$ to be valid such that \textsc{Dijkstra} may have to be adapted accordingly.
This process of finding the shortest path and removing vertices is repeated until paths for all pairs in $C$ are found or no further non-overlapping path can be found. 
It is guaranteed that no overlapping paths can occur since consumed vertices are removed from $\calR$. 
Running this \emph{shortest-first VDP subroutine} yields a set of paths $W_i=\{p_1,...p_n\}$ that correspond to logical two-qubit gates, which can be executed in parallel with lattice surgery on the substrate.

In general, a $\cx{}$ quantum circuit contains multiple sets of pairs $C_i$ as we do not assume that all $\cx{}$ gates have disjoint qubit support. 
Thus a $\cx{}$ quantum circuit can be viewed as $\tilde{C}=\{C_1,...,C_m\}$. 
Additionally, the routing graph $\calR$ may not allow us to find a set of vertex-disjoint paths for a given $C_i$. 
Thus, the aforementioned subroutine must be embedded in an algorithm, which we call \emph{shortest-first VDP solving}.
We start by employing the shortest-first VDP subroutine for $C_1$ and add the resulting set of paths $W_i$ to the overall solution $C'$. 
If not all pairs in $C_1$ can be connected by non-overlapping paths, the remaining elements from $C_1$ are copied to $C_2$.
This may lead to $C_2$ containing gates with shared qubit labels. 
Thus, $C_2$ must be fixed by pushing problematic gates into $C_3$, such that $C_2$ contains gates with disjoint qubit labels only. 
Then, $C_3$ and the subsequent sets may be adapted in the same manner.
Note that pushing requires us to respect the order of gates, as otherwise we could shuffle non-commuting gates.
This shortest-first VDP subroutine together with the pushing is repeated until a path for each pair is found. 
The solution contains multiple sets of paths $C' = \{W_1,...W_\Delta\}$, where $\Delta$ is the depth of the final, compiled circuit.

To include $\tgate$ gates in this picture, one can expand the shortest-first VDP subroutine to allow singleton tuples, i.e., single vertices $r_\ell \in V_\calL$ corresponding to logical qubit labels via $\Lambda^*$. 
Additionally, the vertices of the magic state patches $V_\calF$ must be taken into account in the routing problem formulation.
We model magic state patches as logical patches that ``receive'' magic states after some fixed reset time from some adjacent magic state factory. %
The $\tgate$ gate can be implemented by gate injection from any available magic state patch. 
Thus, one can expand the shortest-first VDP subroutine by greedily computing all paths from the logical data qubit on which a $\tgate$ gate should be applied, $r_\ell$, to the magic state patches from $V_\calF$, if their $\tgate$ state is available. 
The shortest paths between $r_\ell$ and the available magic state patches is compared to all other paths corresponding to elements in $C_i$ and the shortest is added to the solution as before. 
The availability of magic states is modeled by defining a reset time $t\in \mathbb{Z}$, which is initialized as positive integer and
reduced by $1$ after each completion of the shortest-first VDP subroutine. 
The factory is available when $t\leq 0$ and reset to the original value if the magic state was consumed. 
Note that is a pessimistic assumption as the units of time required to create a magic state may be less coarse grained.

The depth of the compiled circuit $\Delta$ depends on the choice of logical qubit labels via $\Lambda^*$, hence it is advantageous to optimize the labeling as well, which we do by using a combinatorial optimization heuristic. 

\subsection{$\Lambda^*$ Optimization with Hill Climbing}

The aim of optimizing the labeling $\Lambda^*$ is to exploit the parallelism of the circuit as this directly influences the shortest-first VDP solving. 
The optimization requires defining a metric that assesses the quality of some choice of $\Lambda^*$.
\begin{example}
    The number of crossings of shortest paths between pairs in $\tilde{C}$ can serve as reasonable cost metric~\cite{molavi_dependency-aware_2024}. 
    Consider~\autoref{fig:fig1}{c}, where an optimal solution can be found by directly computing the shortest paths between the qubit vertices. 
    Swapping the locations of two labels, e.g., $0$ and $3$, however, would inevitably lead to a suboptimal solution because the respective shortest paths would cross. 
    While for this small example the intuition reflects an exact solution, for larger instances counting the crossings of shortest paths will be a heuristic metric in general.
\end{example} 
More formally, this metric can be defined as computing all shortest valid paths in $\tilde{C}$ on $\calR$. %
This yields a list of paths $U_i$ for each $C_i\in\tilde{C}$, i.e., $U = \{U_1,...,U_m\}$.
If a vertex appears in multiple paths it is considered as one or more crossings. 
More specifically, for some $U_i$, the crossings are computed as $c_i = \sum_{v\in V_\calA}\binom{K}{2}$, where $K$ denotes in how many paths $v$ appears. 
Thus each pair of paths sharing a vertex is considered as a crossing. 
Overall, the crossings are counted for each ``layer'' separately, i.e., $c' = \sum_{i=1}^m c_i$.
Note that paths to factories are not considered in this metric, as the availability of a particular factory depends on the reset time until a $\tgate$ state can be consumed from a factory, which is not reflected here but only considered in the routing process discussed above.

To consider the dependency to the factories as well, one could use the final depth $\Delta$ from an execution of the shortest-first VDP solving as cost function, which is more expensive to compute, however.

Hill climbing~\cite{yang_nature-inspired_2014, gendreau_handbook_2019, jacobson_analyzing_2004}, is a simple meta-heuristic that can find local extrema. 
The algorithm uses a random choice of a mapping $\Lambda^*$ as initial value and computes the metric for each so-called neighboring solution.
The algorithm proceeds in a greedy fashion: Once the best solution among the neighboring ones is chosen based on the defined metric, the next neighborhood is searched.
This iterative procedure is repeated until the number of maximal iterations is reached or no better solution in the neighborhood can be found.
We use the simple adaption of random restarts, i.e., running the algorithm with multiple randomly chosen $\Lambda^*$ as initial values. 
This way a slightly larger search space can potentially be explored.
Thus, the number of restarts and the maximal iterations per restart are important parameters of the optimization procedure. 

The neighborhood for the greedy search in the hill climbing is defined as all combinations of swaps of the positions of each pair in $\tilde{C}$~\cite{molavi_dependency-aware_2024}. 
Moreover, the factory positions are not changed during the hill climbing optimization. 
Merely the qubit labels, $\Lambda^*$ are optimized.
Note, however, that one can potentially improve the method by employing more sophisticated algorithms such as ABC~\cite{hancer_artificial_2020} or genetic algorithms~\cite{lambora-genetic}.
Our work is a proof-of-principle demonstration.
We therefore leave a detailed comparison of different optimization strategies open for future work.

\newcommand{\sold}{S_{\text{split}}}
\newcommand{\snew}{S_{\text{merged}}}
\newcommand{\ctg}{\Tilde{\calG}}
\newcommand{\cts}{\Tilde{\calS}}
\newcommand{\cnew}{\calQ_{\text{merged}}}
\newcommand{\cold}{\calQ_{\text{split}}}
\newcommand{\clg}{\calL_g}
\newcommand{\clo}{\calL_\text{split}}
\newcommand{\cln}{\calL_\text{merged}}
\section{Microscopic Substrate realizations
}\label{sec-substrates}
In this section, we outline the subsystem code formalism for lattice surgery introduced by Vuillot et al.~\cite{vuillot_code_2019} and use it to argue fault-tolerance properties of lattice surgery between distant logical patches on a substrate utilizing \emph{distance-preserving snakes}.
We also discuss microscopic realizations of the color code and surface code substrates.
 
\subsection{Lattice Surgery Schemes from Distance-Preserving Snakes}\label{sec:sscode-formalism}

We consider two codes defined by stabilizers on the separate patches and the merged patch $\cold = \langle \sold \rangle, \cnew = \langle \snew \rangle$.
To form a joint subsystem code $\calQ$ we define the gauge group $\calG$ to be generated by $\sold, \snew$, and additional operators that are the anti-commuting conjugate partners of elements in $\calM_p \subseteq \sold \cap \cln$ and $\calM_m \subseteq \snew \cap \clo$.
Adding these conjugate partners ensures that the center of the gauge group contains no elements from $\calM_p, \calM_m$ and thus results in the stabilizer group
$S = Z(\calG) = \sold \cap \snew$.
Finally, the gauge operators acting non-trivially on the gauge qubits are defined as $\calL_g = \calG \setminus S$ and the bare logical operators $\calB = C_\calG(\calG) \setminus \calG$, where $C_\calG(\cdot)$ denotes the centralizer in $\calG$ and the dressed logical operators $\calD = C_\calG(S) \setminus \calG$. 

To argue fault-tolerance properties, we cast lattice surgery between two distant patches connected by a snake on the substrate in this formalism.
In particular, we want to ensure that we can always measure $Z_LZ_L$ or $X_LX_L$ fault tolerantly between two distant logical patches by connecting them via a snake that has a \emph{distance preserving} property. 
We define a distance preserving snake as ancilla region between two logical patches that allows us to implement a fault-tolerant lattice surgery scheme.
In the following we show that we can indeed construct distance preserving snakes for the color code and surface code substrates. 
To this end, we construct a common subsystem code such that the split and merged patches are different gauge fixes of operators of $\calQ$ and examine the corresponding logical operators.

\subsection{Color Code Substrate}
\label{sec-cc-snakes}

\begin{figure}[t]
    \centering
    \begin{subfigure}[b]{0.45\linewidth}
        \centering
        \includegraphics[width=\linewidth]{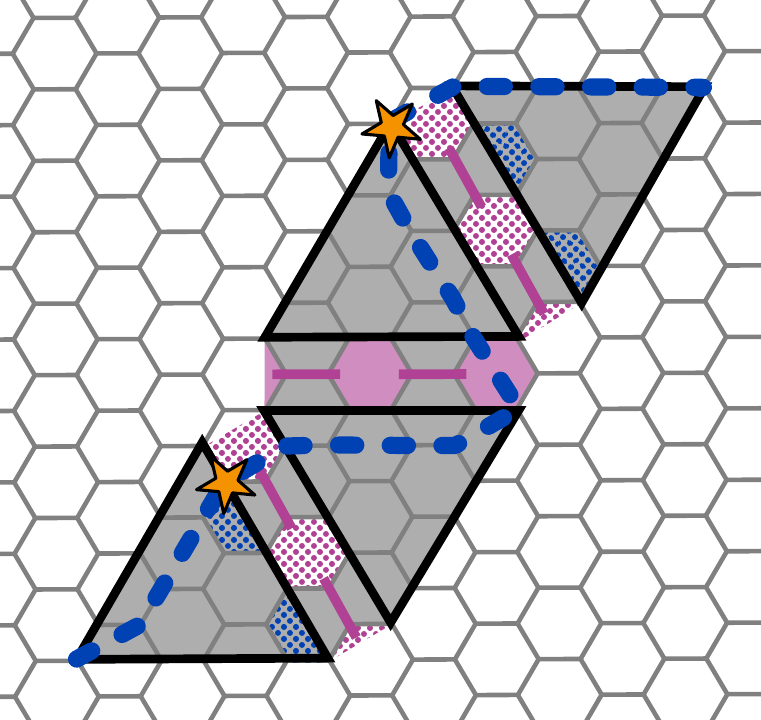}
        \caption{}
        \label{fig:subfig_b_CCSNAKE}
    \end{subfigure}
    \begin{subfigure}[b]{0.45\linewidth}
        \centering
        \includegraphics[width=\linewidth]{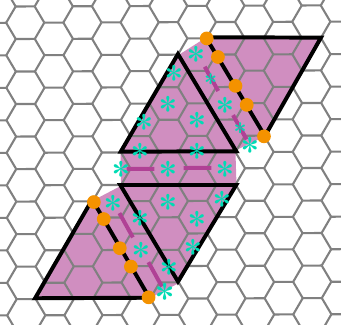}
        \caption{}
        \label{fig:subfig_c_CCSNAKE}
    \end{subfigure}
    \caption{
    Color code snake together with logical ancilla and data qubit. (a) Stabilizers $S$ of the subsystem code $\calQ$.
    Grey plaquettes host $X$ and $Z$ stabilizers and additional $Z$ stabilizers are depicted in solid pink. 
    Elements of $\clg$ correspond to pink and blue dotted faces.
    The blue dashed line is an $X$ bare logical operator; its parts stretching from from the left bottom corner to one of the two orange stars illustrate two possible gauge operators.
    (b) $Z$ stabilizers of the merged code.
    Asterisks mark the subset $M$ of $Z$ stabilizers needed to measure the logical $Z_L Z_L$ operator (supported on the orange circles).
    }
    \label{fig:cc-snake}
\end{figure}
\begin{figure}[t]
    \centering
    \includegraphics[width=0.9\linewidth]{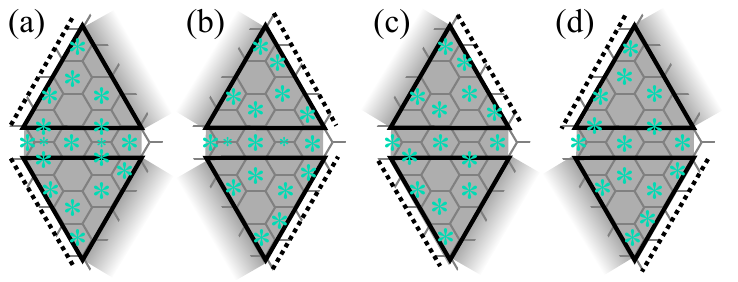}
    \caption{
    Stabilizer subsets marked by asterisks for the logical $X_LX_L/Z_LZ_L$ measurements along the STDW between two ancilla patches of the color code. 
    Outer boundaries without adjacent ancilla patch in the snake are marked by dotted lines; it is required that each qubit on such a boundary is ``touched'' an even number of times by a stabilizer of fixed type. 
    On the other (inner) boundaries where the snake may extend (shaded in gray), qubits are touched only once.
    Depending on how the patches are placed on the substrate, one may reflect the pairs of patches along the vertical axis. 
    Note that this pattern can be geometrically extended to larger distances.} 
    \label{fig:cases-subset-ZZ}
\end{figure}
For the color code substrate we choose a hexagonal tiling on which we define the regions $\calA^*\cup \calL\cup \calF$ that form distance $d=2t+1$ triangular color code patches, where $t$ is a positive integer. 
For this substrate each patch has $q=3t^2 + 3t + 1$ physical data qubits.

To perform lattice surgery between two neighboring logical patches,
their adjacent boundaries are connected by measuring intermediary operators that create a \emph{semitransparent domain wall} (STDW)~\cite{kesselring_anyon_2024}. 
For a $Z_LZ_L$ merge between two adjacent logical patches, one extends the existing $Z$ and $X$ stabilizers and uses further weight $2, 3, 5$, and $6$ stabilizers along the STDW. %
For a $X_LX_L$ merge, the roles of $X$ and $Z$ stabilizers must be swapped. 
For ease of presentation we focus on the case of measuring $Z_L Z_L$.

To argue fault-tolerance properties of lattice surgery on the color code substrate, we define the joint subsystem code $\calQ$ as introduced above to construct distance preserving snakes for $Z_LZ_L$ lattice surgery on the color code substrate.
In the following we focus on the case illustrated in~\autoref{fig:cc-snake}.
$\sold$ is defined by two disjoint logical color code patches and additional disjoint ancilla stabilizers supported on the snake.
$\snew$ is defined on the logical color code patches merged together by a snake whose $X$ stabilizers are the gray faces in~\autoref{fig:cc-snake}{a} and the $Z$ stabilizers are depicted in~\autoref{fig:cc-snake}{b}.

The set of gauge operators $\clg$ contains the new intermediary weight $3,5,$ and $6$ $Z$ plaquettes $\calI$ along the STDWs next to logical patches, depicted as pink dotted plaquettes in ~\autoref{fig:cc-snake}{a}. 
These are the gauge operators that will be fixed for the merge. 
The stabilizers in $\sold$ commute with the new plaquettes, except for the weight 4 $X$ checks along the boundaries of the logical patches, which are also in $\clg$ (blue dotted plaquettes in~\autoref{fig:cc-snake}{a}).
These can be extended to weight 6 stabilizers of $\calQ$ along the STDW. %
To ensure that the stabilizer group of $\calQ$ has the correct form, we add $X$ string operators to $\clg$ as anti-commuting conjugate partners of the operators in $\calI$. 
To do this, it is helpful to distinguish an \emph{inner} boundary of the patch, i.e., a boundary which is connected to another patch with a STDW and an \emph{outer} boundary without adjacent connected patch.
Starting from the left logical patch $L_1$, a string operator can be chosen on $L_1$ such that one of its two point-like excitations is on the outer boundary
of $L_1$, and the other excitation appears on an element from $\calI$ on an inner boundary. 
One can add another string operator extending through the snake that can create an excitation on an element of $\calI$ on the inner boundary adjacent to the top right logical patch (cf.~\autoref{fig:cc-snake}{a}). 
This choice of gauge operators ensures that they commute with the stabilizers of the subsystem code (grey faces) and anti-commute with elements of $\calI$, which are not supposed to be stabilizers, as they anti-commute with the gauge operators of the split code (blue dotted faces).

To fix the gauge, the operators in $\calI$ are measured and the stabilizers are updated according to the Gottesman-Knill theorem~\cite{gottesman1998heisenberg}.

Note that the domain walls dictate a special behaviour of the logical string operators. 
Logical $Z_L$ strings can terminate at the STDW such that the weight of $Z_L$ operators remains the same as for the initial logical patches. 
$X$ strings cannot terminate at the STDW but only on the color code boundaries of the appropriate color~\cite{kesselring_anyon_2024}.
Thus $\cnew$ and $\calQ$ have logical $Z$ operators of weight $d$ of a single logical patch.
Bare logical $X$ operators of $\calQ$ can stretch from one logical patch to another, but since there are $X$ gauge operators supported on the left logical patch and the snake, the minimum distance of a dressed logical operator of $\calQ$ is $d$.
In the example in~\autoref{fig:cc-snake} the resulting operator is supported on the top right logical patch.
Hence, the code distance of the subsystem code is at least the same as of an individual logical patch.

Finally, we have to define a subset $M$ of operators whose measurement outcomes determine the desired $Z_LZ_L$ value.
$M$ must include the operators on the qubits at the inner boundaries
given by $\calI$.
Here, we stick to $Z_L$ along the inner boundaries of the logical patches, but other representatives of $Z_LZ_L$ may be chosen as well.
Since the substrate is a 3-valent graph and the snake slightly differs depending on its exact shape, the definition of $M$ requires special care.

The construction of $M$ starts by choosing suitable stabilizers with support on each ancilla patch on the snake connecting the logical patches.
The outer boundaries on the snake's patches are important as $M$ must include all stabilizers with support on such a boundary to ensure that each qubit is in the support of an even number of stabilizers.
After specifying the stabilizers for all ancilla patches, suitable stabilizers must be chosen along the domain walls connecting inner boundaries. 
As exemplified in~\autoref{fig:cases-subset-ZZ}, the analysis reduces to four possible cases between two neighbouring patches.
The snake and logical patches illustrated in~\autoref{fig:cc-snake}{b} require case (d) between ancilla patches on the snake. 
In this case we have to choose the intermediary stabilizers equivalent to case (b) and (c) for the displayed example. 
Due to the translational invariance of the hexagonal tiling as well as the periodicity of the STDW's structure and the logical patches, we can always find such a pattern for color code patches. %

The situation is equivalent for longer snakes, where the subsystem code is enlarged by additional stabilizers supported on the additional ancilla patches along the snake.

For the split we simply reverse the roles of $\cnew$ and $\cold$, split the previously extended weight 6 stabilizers and apply the Gottesman-Knill rules.

Note that the color code lattice surgery scheme can also be done with $\calA\backslash\calA^*=\emptyset$, as proposed in~\cite{landahl_quantum_2014,kesselring_anyon_2024}, but requires higher weight stabilizers.

Finally, our source code provides a method to construct both the stabilizers of such merged objects as well as the set $M$ for in principle arbitrary distances and number of patches.%

\subsection{Surface Code Substrate}\label{sec:fsc-snakes}
\begin{figure}[t]
    \centering
    \begin{subfigure}[b]{0.4\linewidth}
        \centering
        \includegraphics[width=\linewidth]{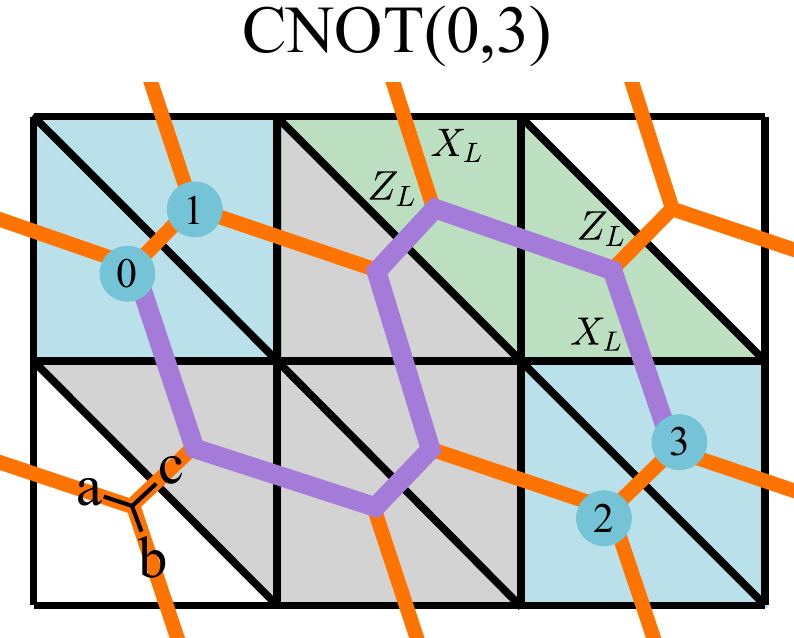}
        \caption{}
        \label{fig:subfig_a_fsc_routing}
    \end{subfigure}
    \begin{subfigure}[b]{0.4\linewidth}
        \centering
        \includegraphics[width=\linewidth]{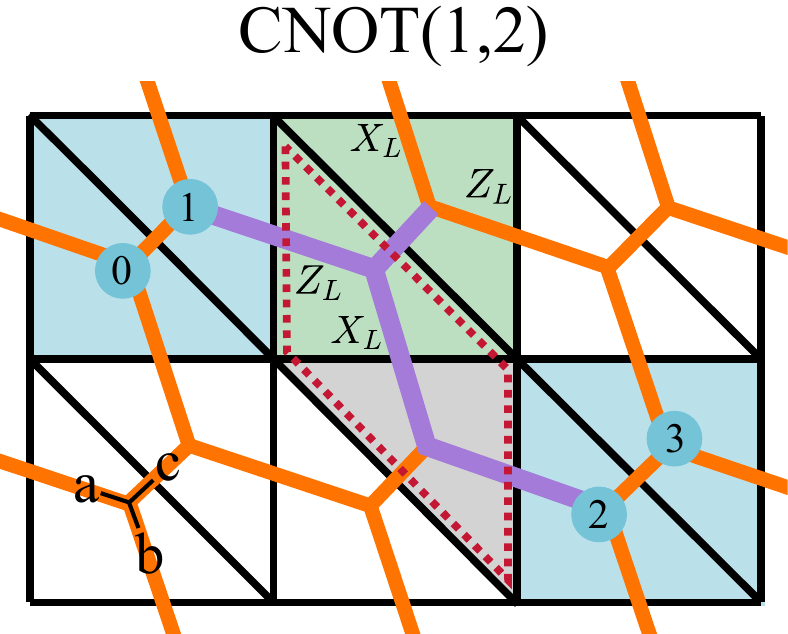}
        \caption{}
        \label{fig:subfig_b_fsc_routing}
    \end{subfigure}
    \caption{Valid paths on the surface code substrate with folded surface codes as logical data patches. 
    Folded surface codes are displayed in light blue, snakes in grey, and logical ancilla patches in green. 
     $\calR$ is shown in orange and a routing path between logical qubits is marked in violet.
     (a) The shortest path on $\calR$ between logical data patches $0$ and $3$ is a valid path. 
     (b) The shortest path on $\calR$ between logical patch 1 and 2 is not valid, since the pair of triangles marked by red dashed lines would allow us to do either $X_LX_L$ or $Z_LZ_L$ lattice surgery, but not both as needed for the implementation of a $\cx$ gate.
     A slightly adapted valid path is displayed in violet together with a possible logical ancilla patch placement.}
    \label{fig:restricted_path_sc}
\end{figure}

As the surface code substrate we consider a square tiling and propose to use the folded surface code to encode logical data qubits.
To ``hide'' the folding diagonal\footnote{To the best of our knowledge lattice surgery along the folding diagonal of the folded surface code is not possible without applying additional local unitaries that in general change the structure of the tiling.},
we place folded surface codes on the substrate in pairs to form squares (cf.\ the pair layout in ~\autoref{fig:layouts_nx}).
For each folded surface code, we consider a stacked layer architecture
to support its two halves or we project the two halves onto one layer with more complex qubit connectivity.
The logical data qubits allow to access both $X_L$ and $Z_L$ for lattice surgery on the short boundaries due to the folded structure.
The logical patches are connected by snakes consisting of a single layer compatible with the standard (not folded) surface code.

Logical ancillas for the realization of $\cx$ gates
comprise two neighboring triangles from $\calA^*$, forming square- or parallelogram-shaped regions. 
Moreover, we assume that the magic state patches are standard surface codes as well.
The snakes forming diagonal parts on this substrate are equivalent to rotated surface codes and the horizontal and vertical parts of a snake are standard surface codes.

The resulting routing graph is hexagonal (cf.~\autoref{fig:overview-figure}), however, since only one type of logical ($X_L$ or $Z_L$) is available per boundary of standard surface code patches, the definition of valid paths for this substrate differs from the color code substrate.
In particular, (i) a valid path must be large enough to contain a logical ancilla patch needed to implement the measurement-based lattice surgery scheme for a $\cx$ and (ii) the path must allow that this ancilla patch can be placed in a way that ensures that both $X_LX_L$ and $Z_LZ_L$ measurements can be performed using the snake.
To see this, assign one of three directions $a,b$, or $c$ to each edge of the hexagonal routing graph.
For a routing path to be valid, it must contain at least one edge of each direction as shown in~\autoref{fig:restricted_path_sc}.
In the considered setting where logical ancillas for the $\cx{}$ surgery are placed on the snake, paths that do not include all types of edges cannot give access to the required boundaries of the logical ancilla.
Hence, connecting two distant logical qubits is %
possible as long as the chosen layout is sufficiently sparse and slightly larger paths that include edges of all directions can be chosen instead of the shortest paths on $\calR$.

\begin{figure}[t]
    \centering
    \includegraphics[width=0.98\linewidth]{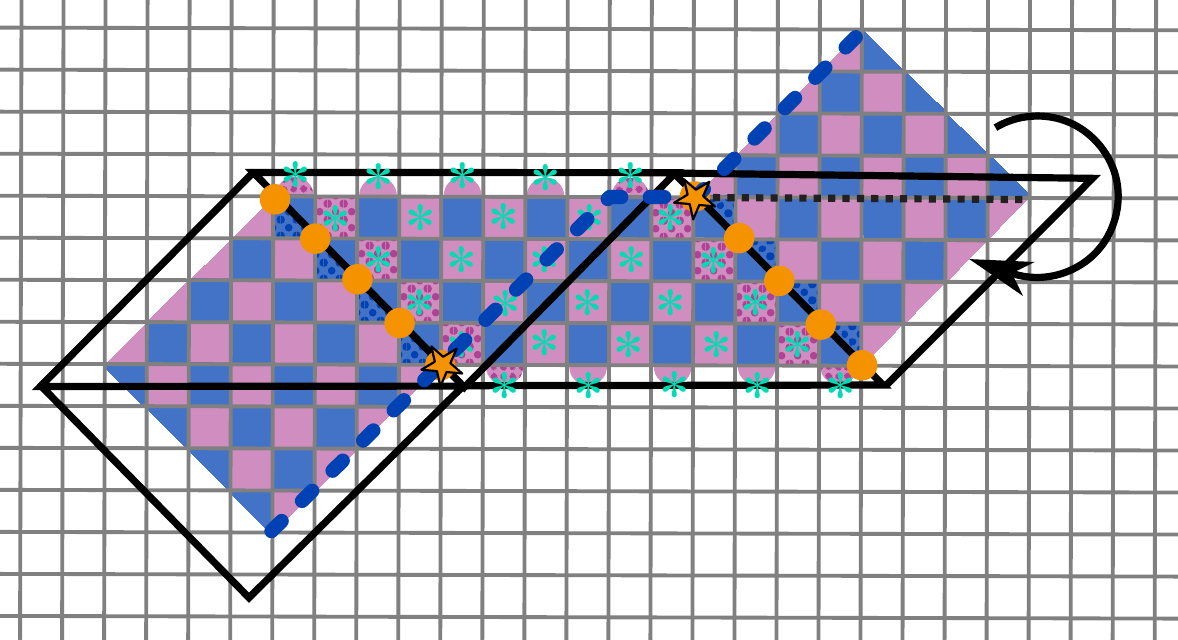}
    \caption{Surface code snake merged with a logical ancilla patch on the left and a folded surface code logical patch on the right. 
    The arrow indicates the fold.
    Each logical patch has distance $d=5$. 
    $X$ stabilizers are displayed in blue and $Z$ stabilizers are displayed in pink. Dotted faces indicate elements in $\clg$.
    The product of $Z$ stabilizers to consider for the $Z_LZ_L$ measurement (supported on qubits with orange dots) are marked with asterisks. An $X$ bare logical and the corresponding gauge operators are shown by the blue dotted line and stars following the same logic as in~\autoref{fig:cc-snake}.}
    \label{fig:fsc-snake}
\end{figure}

Finally, there is an important detail when considering surface code snakes, surface code logical ancillas, and folded surface code logical data patches, as illustrated in~\autoref{fig:fsc-snake}. Considering the left logical ancilla qubit enclosed by the black boundary of two triangles, one can observe that there is in principle some freedom on how one effectively places the logical qubit within the triangles. Imagine the case in which one wants to perform lattice surgery between this logical ancilla patch and another adjacent logical data patch (not shown in the figure). Due to the freedom of effective placement within the black triangles, it may be necessary to define slightly deformed stabilizers between the logical qubits to perform lattice surgery if they are not aligned perfectly, however it can still be implemented with local qubit connectivity.

We illustrate how to construct distance preserving snakes, by considering the surface code snake illustrated in~\autoref{fig:fsc-snake}, which connects a folded surface code patch and a logical surface code patch.
Here, $\sold$ consists of the stabilizers of the disjoint logical patches and stabilizers on the ancilla space on the snake.
$\snew$ consists of one joint patch formed by additional checks in the rotated surface code part of the snake.
The intermediary $Z$ checks $\calI$ on the snake are supported on the weight 2 and 4 $Z$ checks that act on one of the qubits marked with orange dots in~\autoref{fig:fsc-snake}.
These are in $\clg$ and are again the ones we gauge fix for the merge.
Moreover, the weight $3$ $X$ checks along the boundary of the logical ancilla patch and the folded surface code patch anti-commute with the new $Z$ checks formed on plaquettes in $\calI$. 
These are also in $\clg$ and can be extended to weight 4 faces that commute with the new $Z$ checks in $S$. 
Similarly to the color code substrate, one can choose $X$ string operators in $\clg$ with support on the left logical patch and on the snake to be anti-commuting conjugate partners of elements in $\calI$.

The $Z$ distance of $\cnew$ and bare logical operators of $\calQ$ is the same as for $\cold$.
Logical $X$ operators of $\calQ$ stretch from one logical patch to another.
However, these bare logical operators can be multiplied by $X$ gauge operators such that the dressed logical operators are guaranteed to have support on the right logical patch and hence preserve distance $d$.

Finally, to obtain the value of $Z_LZ_L$, we have to define a set $M$ of operators whose measurement outcomes give the value of the operator.
For the surface code substrate these are straightforward to find, due to the checkerboard structure.
Hence, the set $M$ corresponds to the plaquettes marked by the asterisks on the snake in~\autoref{fig:fsc-snake}.
An automatized construction for surface code snakes merged with logical patches is provided in the open-source package.

\section{Numerical Insights Into the Macroscopic Compilation With the Color Code}\label{sec-numerics}

Here we describe numerical experiments to explore how the depth, number of available magic state patches, reset time
of the magic state factories, and the chosen optimization metric influence the behavior of the proposed macroscopic compilation routine from~\autoref{sec-compilation} for the color code substrate. The implementation is part of the \textit{Munich Quantum Toolkit}~\cite{mqt-handbook}.

\subsection{Setup}
The macroscopic compilation operates on the logical routing graph $\calR$.
The valid paths on $\calR$ for the color code substrate do not require special constraints, despite the fact that paths must be large enough to host at least one logical ancilla patch in order to perform a $\cx$ between logical data patches.
This is ensured by the graph structure and the requirement that no direct connections between logical data patches are allowed.
We consider three choices of layouts as illustrated in~\autoref{fig:layouts_nx}. 
Magic state patches are placed at the boundary of the routing graph and connected with two edges to the bulk of the routing graph, while the remaining edge is reserved for the realization of the magic state factory.
We assume that for each factory a new $\mathrm{T}$ state is available after some time interval given as parameter.

\begin{figure}[t]
    \centering
    \includegraphics[width=0.9\linewidth]{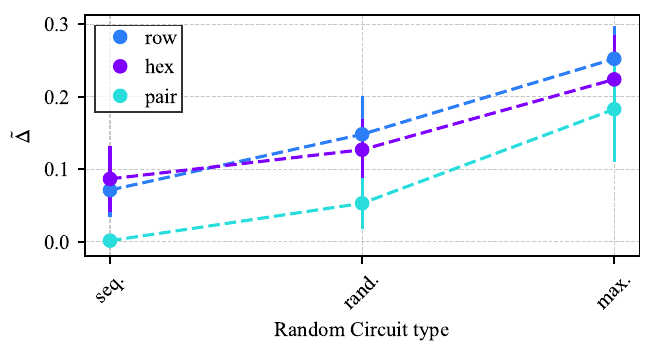}
    \caption{Mean improvements from the hill climbing depending on different circuit types. Here, purely random $\cx{}$ circuits without T gates are considered. ``seq.'' means that at most two $\cx{}$s can be applied in parallel. The ``rand.'' circuits have $\cx{}$s drawn from a random distribution. ``max.'' means that all available qubits can be used in parallel, i.e., $q/2$ gates are applied in parallel. Per data point, 50 random circuits with $q=24$ qubits were sampled. Each run of the hill climbing algorithm uses 50 iterations for each of the 10 restarts. In total, $4q$ gates per circuit were used.}
    \label{fig:parallelism_plot}
\end{figure}

\subsection{$\cx{}$ Circuit Compilation}
Naturally the amount of possible parallelization of the input circuit $\tilde{C}$ has direct impact on the possible depth of the compiled circuit $C'$.
In the language of the routing problem from~\autoref{sec:shortest-first-vdp}, the parallelism of $\tilde{C}$ is higher if it needs to be decomposed into fewer subsets $C_i$ for a fixed number of total gates. 
This initial parallelism may be increased in principle by (pre)-compilation on the circuit-level.
The initial parallelism does not only affect the shortest-first VDP solving, but also the potential of improving $\Lambda^*$. 
To study the behaviour of the hill climbing optimization for different levels of initial parallelism while ignoring effects from $\tgate$ gates, we consider three types of uniformly random $\cx{}$ circuits. 
First, the ``sequential'' (seq.) type constructs circuits in such a way that only two gates can be applied in parallel in each circuit layer in principle as only two subsequent gates have disjoint qubit labels.
The ``random'' (rand.) circuit type directly samples the target and control labels of gates from a uniform distribution.
The ``maximal'' (max.) type has maximal initial parallelism in the sense that $q/2$ subsequent $\cx{}$ gates have disjoint qubit labels for $q$ logical qubits.

The mean improvements for the hill climbing for these randomly sampled circuits for different layouts are displayed in~\autoref{fig:parallelism_plot}. 
The improvement is defined as $\tilde{\Delta}=(\Delta_i - \Delta_f)/\Delta_f$ where $\Delta_f$ is the depth of $C'$ computed with the best $\Lambda^*$ given the heuristic crossing metric. 
In turn, $\Delta_i$ is the depth of $C'$ with the corresponding random initialization of $\Lambda^*$. 

One can observe the tendency that circuits with a higher degree of parallelism can indeed be improved more by the hill climbing optimization with the crossing metric.
Note that the error bars have considerable size because randomness does not only enter for the randomly sampled circuits but also due to the randomly sampled restart instances $\Lambda^*$ for the hill climbing. 

The comparison between different layouts shows that the optimized $\Lambda^*$ on the pair layout tends to be improved less.
A reason for this may be that sparser layouts with larger routing space encounter less crossings during the VDP solving such that one can assume less potential of $\Lambda^*$ improvement for them. 
However, if the initial parallelism is large enough---increasing the possibility of crossings---also optimization for the pair layout becomes more fruitful.
Overall, $\tilde{\Delta}$ increases with growing parallelism for all layouts. The asymptotic packing ratio of the row and hexagonal layout are very close which is why they show similar behavior.
\begin{figure}[t]
    \centering
    \includegraphics[width=0.8\linewidth]{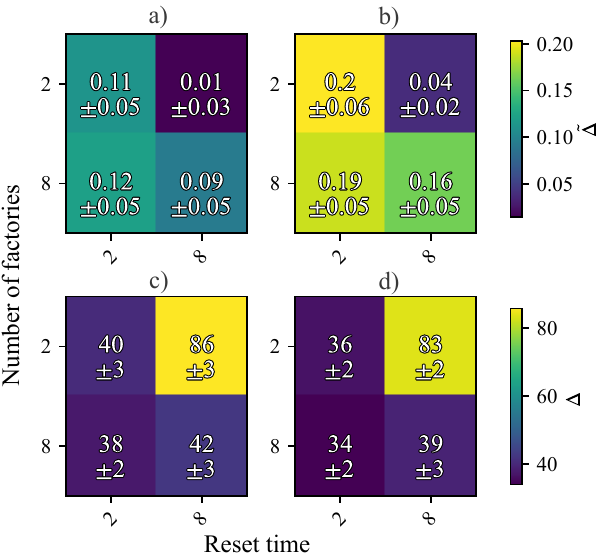}

    \caption{Mean improvements $\tilde{\Delta} = (\Delta_i - \Delta_f) / \Delta_i$ for the heuristic crossing metric (a) and the exact routing metric (b). Absolute depth $\Delta_f$ after optimization for the heuristic crossing metric (c) and the exact routing metric (d). Per data point 50 different ``random'' circuits with $q=24$ qubits and $r=0.8$ were sampled. The ``row'' layout was chosen as well as the $\calR$ for the color code substrate. Per circuit, $4q$ gates were considered.}
    \label{fig:q42_hc_abs}
\end{figure}

\subsection{$\cx{}+\tgate$ Circuit Compilation}
Here we consider $\tgate$ gates in the input circuits as well.
To this end, random circuits are sampled in such a way that there will be a specific ratio of $\cx{}$ gates in comparison to the total gates $r=\# CNOT / (\#CNOT +\#T)$. 
This ratio is important as $\tgate$ gates make an efficient solution particularly difficult due to the reset time $t$ and the limited availability of magic state patches.
We investigate the behavior of the hill climbing optimization with different numbers of magic state patches as well as different reset times. The results for ``random'' circuits are shown in~\autoref{fig:q42_hc_abs}. 
The first row displays the improvements $\tilde{\Delta}$ for the heuristic crossing metric (a) and the more expensive metric $\Delta_f$ of the shortest-first VDP solving (b).
The improvements work best for many magic state patches and a small reset time, while they are most likely reduced for more challenging setups. 
Using the exact metric $\Delta_f$ makes the hill climbing solutions more advantageous than with the heuristic.
The bottom row of~\autoref{fig:q42_hc_abs} displays the corresponding absolute values for $\Delta_f$ of the VDP solving runs of the best $\Lambda^*$ from the hill climbing optimization for the respective metrics (c, d). 
It is apparent that irrespective the chosen metric the achieved depth is the lowest for many magic state patches and low reset time $t$, with slightly better results using the metric $\Delta_f$ for hill climbing.

\section{Conclusion and Outlook}\label{sec-conclusion}
Our work presents a generalized approach to fault-tolerant compilation based on lattice surgery, independent of any specific topological code.
We split the overall compilation into two levels of abstraction, microscopic and macroscopic, analyzing two substrate choices and proposing a routing and mapping routine.
We investigated how microscopic design choices influence the macroscopic compilation task and provided initial numerical results for the color code, highlighting how logical circuit parallelism and magic state availability affect overall performance.

Our methods enable future optimizations of fault-tolerant compilation.
At the microscopic level, we envision exploring different substrates, possibly motivated by different quantum computing platforms, and taking hardware imperfections into account.
For instance, one possible direction is to incorporate into our framework the recently-developed methods for adapting surface code patches in the presence of defective qubits and gates~\cite{auger2017fault,debroy_luci_2024,leroux_snakes_2024}.
At the macroscopic level, the $\Lambda^*$ optimization and algorithms to solve the (extended) VDP problem allow for further improvements; commuting some Clifford gates to the end of the circuit and exploring potential parallelism in the modified circuit may be valuable.  
Future work may also explore how to incorporate novel schemes for magic state distillation and cultivation~\cite{gidney_magic_2024}, with a careful analysis of their layouts within the bulk of the routing graph.

\section*{Acknowledgements}
L.H., L.B., and R.W. acknowledge funding from the European Research Council (ERC) under the European Union’s Horizon 2020 research and innovation program (grant agreement No. 101001318). 
This work was part of the Munich Quantum Valley, which the Bavarian state government supports with funds from the Hightech Agenda Bayern Plus as well as by the BMK, BMDW, and the State of Upper Austria in the frame of the COMET program (managed by the FFG).

\clearpage
\balance
\printbibliography

\end{document}